\documentclass[doublecol]{epl2} 
\usepackage{graphicx,amssymb,amsmath}
\usepackage[colorlinks,bookmarks=false,citecolor=blue,linkcolor=red,urlcolor=blue]{hyperref}
\usepackage{color}

\newlength{\ldag}
\begin{document}

\title{Effective models for gapped phases of strongly correlated quantum lattice models}

\author{H. Y. Yang\inst{1} \and K. P. Schmidt\inst{1}}
\shortauthor{H.Y. Yang and K. P. Schmidt}

\institute{
\inst{1} Lehrstuhl f\"ur Theoretische Physik I, Otto-Hahn-Stra\ss e 4, TU Dortmund, D-44221 Dortmund, Germany
}



\abstract{
 We present a robust scheme to derive effective models non-perturbatively for quantum lattice models when at least one degree of freedom is gapped. A combination of graph theory and the method of continuous unitary transformations (gCUTs) is shown to efficiently capture all zero-temperature fluctuations in a controlled spatial range. The gCUT can be used either for effective quasi-particle descriptions or for effective low-energy descriptions in case of infinitely degenerate subspaces. We illustrate the method for 1d and 2d lattice models yielding convincing results in the thermodynamic limit. We find that the recently discovered spin liquid in the Hubbard model on the honeycomb lattice lies outside the perturbative strong-coupling regime. Various extensions and perspectives of the gCUT are discussed.     
}

\maketitle
%
%
\section{Introduction}
%
%
A microscopic description of strongly correlated quantum many-body systems is very demanding and powerful theoretical techniques are clearly needed. Especially in two dimensions there is a lack of theoretical tools to tackle microscopic models. Despite quantum Monte Carlo, functional renormalization group, and 2d extensions of density matrix renormalization group, the method of continuous unitary transformations (CUTs) \cite{Wegner94,Glazek93,Glazek94,Kehrein06} has been applied successfully to a variety of physical systems. Here a CUT is used to derive effective low-energy models with the major task to solve an infinite hierarchy of differential equations. One controlled way to proceed is to truncate these equations in a perturbative manner yielding a high-order series expansion of the effective low-energy model \cite{Knetter00,Knetter03_1,Dusuel10}. Such a perturbative version (pCUTs) is similar in spirit to linked cluster expansions (LCEs) which have become a standard tool to treat a large class of quantum lattice models \cite{Gelfand00,Oitmaa06}. LCEs exploit the linked-cluster theorem by calculating on finite graphs the correct amplitudes of the effective low-energy model in the thermodynamic limit up to high orders in perturbation. In contrast to conventional LCEs, the pCUT has been also applied succesfully to infinitely degenerated subspaces \cite{Schmidt08,Dorier08,Yang10} and to topologically ordered ground states \cite{Vidal09_1}.     

The applicability of pCUTs is clearly limited by its perturbative nature. A non-perturbative extension is therefore highly desirable. Attemps in the past solved the flow equations in a self-similar way (sCUTs) by truncating in terms of the spatial extension and the structure of the operators \cite{Dusuel04,Reischl04,Fischer10,Hamerla10}. The major complication with these approaches is to control the error induced by truncations.

Here we present a graph-theory based continuous unitary transformation (gCUT) which yields a robust scheme to derive effective models non-perturbatively by integrating out gapped degrees of freedom in a controlled spatial range. It can be applied for all kinds of lattice topologies, for all kinds of interacting degrees of freedom, and for deriving low-energy models with a reduced Hilbert space.    
%
%
%
%

Let $\hat{H}$ be an Hamiltonian on a given arbitrary lattice at zero temperature. On each vertex of the lattice a quantum degree of freedom like a spin, a boson, or a fermion is placed which interacts via short-range operators. 

We use a CUT to map $\hat{H}$ unitarily to an effective Hamiltonian $\hat{H}_{\rm eff}$. In most cases we focus on a quasi-particle (QP) conserving CUT \cite{Knetter00,Knetter03_1}, i.e. the effective Hamiltonian is blockdiagonal in a QP counting operator $\hat{Q}$. Formally, we can write
$\hat{H}_{\rm eff} = \sum_{q=0}^\infty \hat{H}_{\rm eff}^{(q)}$,
where $q$ denotes the number of QPs. The low-energy physics is then expected to be contained in the blocks with $q$ small simplifying the many-body problem tremendously.

Each QP block of the effective Hamiltonian $\hat{H}_{\rm eff}^{(q)} = \sum_i a_{i,{\rm eff}}^{(q)} \mathcal{O}_i^{(q)}$ is generically a sum of $q$-particle operators $\mathcal{O}_i^{(q)}$ weighted by an amplitude $a_{i,{\rm eff}}^{(q)}$. The final goal of the gCUT is to calculate these amplitudes quantitatively in the thermodynamic limit. The gCUT is defined by the following steps:

(i) Generate all topologically distinct connected graphs $\mathcal{G}_\nu$ on the lattice and sort them by their number of sites $n$. 

(ii) For each graph $\mathcal{G}_\nu$ one sets up the finite number of flow equations using the method of CUTs \cite{Wegner94,Glazek93,Glazek94,Kehrein06}. A continuous auxiliary variable l is introduced defining the l-dependent Hamiltonian $\hat{H}^{{\mathcal{G}}_\nu} (l)$:=$U^\dagger (l)\hat{H}^{{\mathcal{G}}_\nu}U(l)$. Then the flow equation is given by $
 \partial_l \hat{H}^{{\mathcal{G}}_\nu}(l) = [\hat{\eta} (l), \hat{H}^{{\mathcal{G}}_\nu}(l)]$,
where $\hat{\eta} (l)$:=$-U^\dagger(l)(\partial_l U(l))$ is the anti-Hermitian generator of the unitary transformation. At the end of the flow one obtains an effective graph-dependent Hamiltonian $\hat{H}_{\rm eff}^{{\mathcal{G}}_\nu}$ with amplitudes $a^{{\mathcal{G}}_\nu}_{\rm eff}$. For the QP conserving generator, $\hat{H}_{\rm eff}^{{\mathcal{G}}_\nu}$ commutes with a chosen QP counting operator $\hat{Q}$ and the amplitudes $a^{{\mathcal{G}}_\nu}_{\rm eff}$ can be sorted by the number of QPs involved. In an eigenbasis of the operator $\hat{Q}$, the QP conserving generator is given by 
$
 \hat{\eta}^{\rm qp}_{i,j} (l) = {\rm sgn}\left( q_i-q_j\right) \hat{H}_{i,j} (l)$,
where $q_i$ and $q_j$ are eigenvalues of $\hat{Q}$. One can also use the generator introduced by Wegner \cite{Wegner94} which reads
$
 \hat{\eta}^{\rm Wegner}_{i,j} (l) = \left[  \hat{H}^{\rm d}_{i,j} (l) ,  \hat{H}^{\rm nd}_{i,j} (l) \right]$,
where the diagonal part $\hat{H}^{\rm d}$ is given by all matrix elements between states with the same number of QPs and $\hat{H}^{\rm nd}$ denote all remaining non-diagonal processes. Since no truncations are done for a given graph, the flow equations converge \cite{Wegner94,Mielke98,Dusuel04} and, consequently, the gCUT is intrinsically robust. 

(iii) The graph-dependant amplitudes $a^{{\mathcal{G}}_\nu}_{\rm eff}$ contain contributions from smaller subgraphs $\mathcal{G}_\mu$ with $\mu < \nu$ which have to be substracted in order to avoid double counting. We therefore define reduced amplitudes $a_{\rm eff}^{(q)} (\nu)$ representing {\it all} fluctuations where the whole graph with {\it all} $n$ sites are involved
\begin{equation}
 a_{\rm eff}^{(q)} (\nu) = a^{{\mathcal{G}}_\nu}_{\rm eff} - \sum_{\mathcal{G}_\mu\in\mathcal{G}_\nu \,  , \, \mu < \nu} a_{\rm eff}^{(q)} (\mu) \quad .
\end{equation}
The next step is to collect properly all the reduced contributions from the different graphs in order to get the amplitudes $a_{\rm eff}^{(q)}$ in the thermodynamic limit. One finds
\begin{equation}
 a_{\rm eff}^{(q)} = \sum_{n=1}^{\infty} a_{\rm eff}^{(q)} (n) = \sum_{n=1}^{\infty} \sum_{|\nu|=n}^{\phantom{\infty}} m (\nu) \,\, a_{\rm eff}^{(q)} (\nu) \quad .
\end{equation}
Here $m (\nu)$ denotes the integer number how many times the process belonging to the amplitude $a_{\rm eff}^{(q)}(\nu )$ on graph $\nu$ can be embedded on the real lattice.

The demand of the full gCUT calculation scales mainly with the maximal number of sites $n_{\rm max}$ of the largest graphs and their total number. The number of flow equations to solve is given by the square of the symmetry-reduced Hilbert space on a given graph. The number $n_{\rm max}$ corresponds physically to the spatial range of fluctuations treated in the gCUT calculation. So as long as the correlation length is finite which is true for any gapped system, the gCUT enables a quantitative and non-perturbative derivation of effective low-energy models when $n_{\rm max}$ is chosen large enough.
 
%
%
%
%
Before discussing explicitly applications of the gCUT, we present some generic and important properties. In general, one should distinguish two types of applications: i) QPs above a unique ground state and ii) effective low-energy models for infinitely degenerated subspaces.

In the first case, one is interested in the QP properties above a strongly correlated ground state. Here the ground state is up to topological degeneracies unique and it can be characterized by the ground state energy per site which is given by the $q=0$ block of $\hat{H}_{\rm eff}$. The one-particle dispersion can be obtained from the $q=1$ block and, consequently, multi-particle properties are contained in blocks with $q>1$.

Let us mention that for the single-valued $q=0$ block the gCUT just diagonalizes the problem and the calculation then corresponds to exact diagonalization of the finite graphs. Such an exact linked cluster expansion for the ground state energy has been already discovered and used successfully by Irving and Hamer in 1984 in the context of lattice gauge theories \cite{Irving84}, and it has been extended to thermodynamic quantities for quantum lattice models recently \cite{Rigol06}. But the gCUT goes well beyond these techniques for $q>0$.

In the second case, the system consists of different degrees of freedom and only some of them are gapped. Then one is interested in deriving an effective model in the reduced but still infinitely large $q=0$ Hilbert space. One typical example which we will detail below is the strong coupling limit of single-band Hubbard models where an effective description in terms of interacting quantum spins is appropriate since charge degrees of freedom are gapped. 

It is important to understand the difference of the generators $\eta^{\rm qp}$ and $\eta^{\rm Wegner}$. Let us distinguish for both generators the states of the full Hilbert space on a certain graph according to the unperturbed problem: We define the lowest energy states to have $q=0$ and all the remaining states to have $q>0$. One then has to distinguish two regimes: 
a) If all energies corresponding to the unitarily transformed $q=0$ states are energetically below the rest of the states, then both generators give successfully an effective low-energy description. The effective models obtained by the two generators can differ by a unitary transformation of the $q=0$ Hilbert space but yield the exact spectrum of the considered graph.
b) In contrast, if the spectrum of a graph is such that $q=0$ states and $q>0$ states overlap energetically, then the two generators behave differently. It is well known for $\eta^{\rm qp}$ \cite{Knetter00} that it sorts the eigenvalues in ascending order. Then $q=0$ states with a high-energy are replaced by low-energy $q>0$ states in the low-energy block during the solution of the flow equation. As a consequence, no well-defined effective low-energy theory can be derived with $\eta^{\rm qp}$. This is different for $\eta^{\rm Wegner}$ which always keeps track of the unitarily transformed $q=0$ states and one is able to deduce an effective low-energy model. The different behaviour of both generators is explicitely found for the Hubbard model on the honeycomb lattice (see below). 

 The gCUT for $q=0$ is similar to the contractor renormalization group (CORE) \cite{Capponi04}. Indeed, in CORE the exact low-energy spectra of graphs for the full model are matched to the low-energy spectra of an effective Hamiltonian in a reduced Hilbert space. The obtained effective model is identical to the one obtained by gCUT up to a unitary transformation inside the low-energy Hilbert space. Interestingly, also in CORE one has to be very careful if states classified as low- and high-energy states in a perturbative limit overlap energetically \cite{Capponi04}. If one just chooses the states having the lowest energy on a given cluster, then no well-defined effective model can be found like with $\eta^{\rm qp}$. So also in CORE one has to find a way to choose the relevant states in order to build an effective low-energy model \cite{Capponi04}. This is actually done automatically in the gCUT approach when using $\eta^{\rm Wegner}$.

The gCUT can be implemented for any generator $\hat{\eta}$ which is optimal for a given physical problem. If one is for example only interested in certain QP blocks or would like to treat the decay of QPs, it might be advantageous to only partially block diagonalizing the problem. This is exactly achieved with the generators proposed recently \cite{Fischer10,Hamerla10}. 

%
%
%
%
As a first example for type i), we apply the gCUT for the 1d Ising model in a transverse field which is given by
\begin{equation}
 \hat{H}_{\rm TFIM} = -\sum_{i} \left( \sigma^z_i + J \sigma^x_i \sigma^x_{i+1} \right)  
\end{equation} 
with $J$ assumed to be positive. This model is exactly solvable by fermionization \cite{Pfeuty70} and it allows therefore a quantitative gauge of the gCUT. Additionally, it nicely serves as a pedagogical example where the elementary properties of the gCUT can be demonstrated explicitely. The model is self-dual and one has a quantum phase transition at $J=1$. It is therefore sufficient to concentrate on the parameter space $J\leq 1$. The unique ground state is unitarily connected to the polarized state where all spins point in $z$-direction and elementary excitations are dressed spin flips. Consequently, the operator $\hat{Q}$ is chosen to be a counting operator of down spins.   

\begin{figure}
\begin{center}
\includegraphics*[width=0.9\columnwidth]{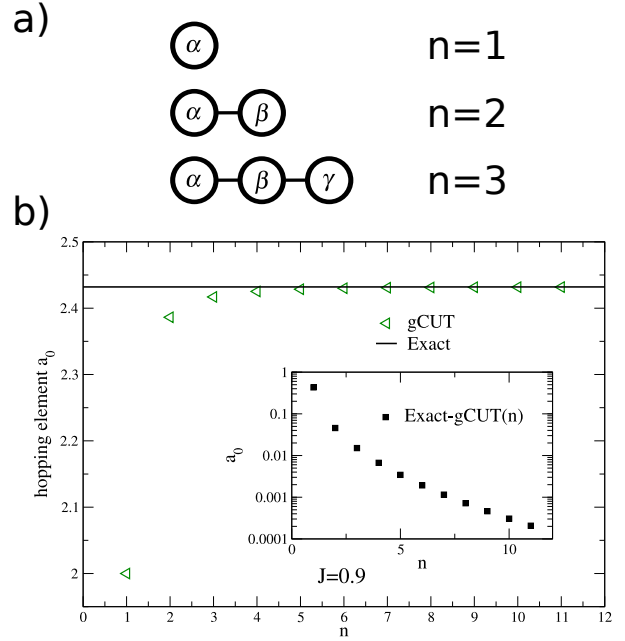}
\end{center}
\caption{(Color online) (a) Illustration of the first three graphs with $n\in\{1,2,3\}$ for the 1d Ising model in a transverse field. The sites on the graphs are denoted by $\alpha$, $\beta$, $\gamma$, et cetera from left to right. (b) The local hopping amplitude $a_0$ as a functions of the graph size $n$ is displayed for $J=0.9$. The gCUT results are denoted by triangles and the exact value is indicated by the solid line. The inset shows the difference between the exact value and the gCUT results on a logarithmic scale.} 
\label{fig:ising1d_illustration}
\end{figure}

The graph theory aspects are particularly simple in 1d. Graphs are chain segments and there exist only one graph for a given $n$ (see Fig.~\ref{fig:ising1d_illustration}a). All graphs $\nu$ can be embedded $N$-times on the lattice and one has $m (\nu)=N$. We have calculated the 0QP and 1QP sector up to $n_{\rm max}=11$. 

%
%
%

%
%
%
\begin{figure}
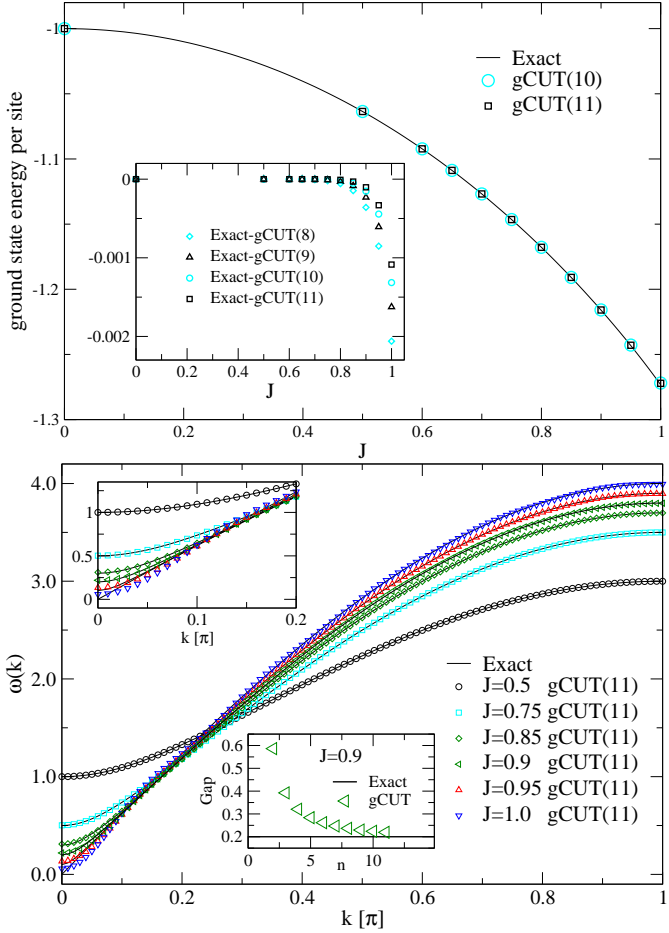

\begin{center}
\includegraphics*[width=\columnwidth]{./fig1.eps}
\includegraphics*[width=\columnwidth]{./fig2.eps}
\end{center}
\caption{(Color online) Comparison of exact and gCUT results for the 1d Ising model in a transverse field: (a) ground state energy per site as a function of $J$. The inset displays differences between exact and gCUT results for various $n_{\rm max}$. (b) One-particle dispersion $\omega (k)$ for different values of $J$. Solid lines represent the exact results while symbols correspond to gCUT results. Upper inset represents a zoom. Lower inset shows the one-particle gap $\Delta=\omega (0)$ at $J=0.9$.} 
\label{fig:ising1d}
\end{figure}

The $q=0$ sector is a single number which corresponds to the ground state energy per site $\epsilon/N$. Following the gCUT recipe, one has for a chosen $n_{\rm max}$
\begin{eqnarray}
 \frac{\epsilon (n_{\rm max})}{N} &=& \sum_{\nu =1}^{n_{\rm max}} \epsilon_{\rm eff}^{(0)} (\nu ) \nonumber\\
                                &=&  \sum_{\nu =1}^{n_{\rm max}} \left( E_{\rm eff}^{\mathcal{G}_\nu} - \sum_{\mu < \nu} \epsilon_{\rm eff}^{(0)} (\mu ) \right) \,  ,
\end{eqnarray}
where $E_{\rm eff}^{\mathcal{G}_\nu}$ is the ground state energy of graph $\nu$ obtained by the gCUT and $\epsilon_{\rm eff}^{(0)} (\nu )$ is the corrected energy where contributions from all subgraphs are substracted. For $n_{\rm max}\leq 3$ one finds the following expressions 
\begin{eqnarray}
 \epsilon_{\rm eff}^{(0)} (\nu = 1) &=& E_{\rm eff}^{\mathcal{G}_1}\nonumber\\
 \epsilon_{\rm eff}^{(0)} (\nu = 2) &=& E_{\rm eff}^{\mathcal{G}_2}- 2 \epsilon_{\rm eff}^{(0)} (1) \nonumber\\
 \epsilon_{\rm eff}^{(0)} (\nu = 3) &=& E_{\rm eff}^{\mathcal{G}_3}- 2 \epsilon_{\rm eff}^{(0)} (2) - 3 \epsilon_{\rm eff}^{(0)} (1)\nonumber
\end{eqnarray}
The corresponding results up to $n_{\rm max}=11$ are shown in the upper panel of Fig.~\ref{fig:ising1d}.  The gCUT results monotonically improve for increasing $n_{\rm max}$. As expected, the largest deviations are found at the critical point $J=1$. But for $J<1$, the gCUT converges quickly to the exact result.   

The $q=1$ sector contains the one-particle dynamics in the form of hopping elements $a_{\delta}$. The amplitude $a_{\delta}$ corresponds to a hopping from site $i$ to site $i+\delta$ on the infinite lattice. A gCUT calculation for a given $n_{\rm max}$ includes therefore hopping elements $a_{\delta}$ with $\delta\in\{0,\pm 1,\ldots,\pm (n_{\rm max}-1)\}$ and one finds  
\begin{eqnarray}
 a_\delta (n_{\rm max}) &=& \sum_{\nu =1}^{n_{\rm max}} a_{\delta, {\rm eff}}^{(1)} (\nu ) = \sum_{\nu =1}^{n_{\rm max}}\sum_{i=1}^{n_{\rm max}-\delta} a_{i,i+\delta,{\rm eff}}^{(1)} (\nu )\\
                                &=& \sum_{\nu =1}^{n_{\rm max}} \sum_{i=1}^{n_{\rm max}-\delta} \left( a_{i,i+\delta,{\rm eff}}^{\mathcal{G}_\nu} - \sum_{\mu < \nu} a_{i,i+\delta,{\rm eff}}^{(1)} (\mu ) \right) \,  ,\nonumber
\end{eqnarray}
where $a_{i,i+\delta,{\rm eff}}^{(1)} (\nu )$ denotes the effective hopping amplitude from site $i$ to site $i+\delta$ on the graph $\nu$ where all subgraph contributions are already substracted.

We would like to illustrate this procedure for the local hopping $a_0$ and for $n_{\rm max}\leq 3$. We denote the sites on the graphs by $\alpha$, $\beta$, $\gamma$ et cetera from left to right (see Fig.~\ref{fig:ising1d}a). The case $n_{\rm max}=1$ is then simply given by
\begin{equation}
 a_{\alpha,\alpha,{\rm eff}}^{(1)} (1) = a_{\alpha,\alpha,{\rm eff}}^{\mathcal{G}_1} \quad . \nonumber 
\end{equation}
For $n_{\rm max}=2$ one finds for the two amplitudes the following expressions
\begin{eqnarray}
 a_{\alpha,\alpha,{\rm eff}}^{(1)} (2) &=& a_{\alpha,\alpha,{\rm eff}}^{\mathcal{G}_2}-a_{\alpha,\alpha,{\rm eff}}^{(1)} (1)\nonumber\\
a_{\beta,\beta,{\rm eff}}^{(1)} (2) &=& a_{\beta,\beta,{\rm eff}}^{\mathcal{G}_2}-a_{\alpha,\alpha,{\rm eff}}^{(1)} (1)\nonumber \quad .
\end{eqnarray}
In practise, we stress that there are of course symmetries of the problem on a given graph which should be considered in order to reduce the effort of the calculation. In the present case all chain segments have a reflection symmetry with respect to the geometrical center of the chain. This immediately implies $a_{\alpha,\alpha,{\rm eff}}^{(1)} (2) = a_{\beta,\beta,{\rm eff}}^{(1)} (2)$. Finally, the analogue formulas for the three processes on the graph with three sites read
\begin{eqnarray}
 a_{\alpha,\alpha,{\rm eff}}^{(1)} (3) &=& a_{\alpha,\alpha,{\rm eff}}^{\mathcal{G}_3} - a_{\alpha,\alpha,{\rm eff}}^{(1)} (2)-a_{\alpha,\alpha,{\rm eff}}^{(1)} (1)\nonumber\\
 a_{\beta,\beta,{\rm eff}}^{(1)} (3) &=& a_{\beta,\beta,{\rm eff}}^{\mathcal{G}_3} \nonumber\\
                               &&- a_{\alpha,\alpha,{\rm eff}}^{(1)} (2)- a_{\beta,\beta,{\rm eff}}^{(1)} (2)-a_{\alpha,\alpha,{\rm eff}}^{(1)} (1) \nonumber\\
a_{\gamma,\gamma,{\rm eff}}^{(1)} (3) &=& a_{\gamma,\gamma,{\rm eff}}^{\mathcal{G}_3} - a_{\beta,\beta,{\rm eff}}^{(1)} (2)-a_{\alpha,\alpha,{\rm eff}}^{(1)} (1)\nonumber \quad .
\end{eqnarray}
The corresponding results for the local hopping element $a_0$ for graphs up to $n_{\rm max}= 11$ are displayed in Fig.~\ref{fig:ising1d}b and compared to the exactly known value.

The one-particle dispersion $\omega (k)$ is obtained by Fourier transformation 
$\omega (k) = a_0 + 2\sum_{\delta=1}^\infty a_\delta \cos (\delta k)$.
 The gCUT results are displayed in the lower panel of Fig.~\ref{fig:ising1d}. The gCUT can again be compared with the analytically known exact result $2\sqrt{1+J^2-2 J\cos(k)}$. The only sizable deviations occur close to the critical point $J=1$ and close to $k=0$. Otherwise, the gCUT one-particle dispersion is basically converged and captures all relevant aspects of the exact expression.
 
%
%
%
%
%

As an example of type ii), we study the Mott phase of the single-band Hubbard model on the triangular and the honeycomb lattice at half filling. The model is given by
\begin{equation}
\hat{H} = U\sum_{i}n_{i\uparrow}n_{i\downarrow} -t\sum_{\langle i,j\rangle,\sigma}(c_{i\sigma}^{\dagger}c_{j\sigma}+\text{h.c.})\quad .
\label{Hubbard_model}
\end{equation} 
In the strong coupling limit $t/U\rightarrow 0$, the system has a finite charge gap and it is expected that the physics can be described solely by spin degrees of freedom \cite{MacDonald88,MacDonald88b}. Indeed, leading second order perturbation theory gives a nearest-neighbor Heisenberg model. But it is an interesting and a challenging problem whether one can derive an effective spin model also close to the metal-insulator transition when the Mott insulator is already weak and corrections to the Heisenberg exchange become important. Physically, interesting spin liquid (SL) phases are expected to be realized in this regime \cite{Morita02,Motrunich05,Meng10,Yang10}. 

The operator $\hat{Q}$ is chosen as a counting operator of double occupancies. In contrast to the previous example where the main interest was in blocks with a finite number of QPs, here we focus on the block with no double occupancies which can be expressed as a spin 1/2 model 
 $H_{\rm eff}^{0} = \sum_{\vec{i},\vec{n}} J_{\vec{n}} \left(\vec{S}_{\,\vec{i}} \cdot\vec{S}_{\,\vec{i}+\vec{n}}\right) + \ldots$,
where $\vec{i}$ and $i+\vec{n}_{\alpha}$ denote sites on the lattice. All remaining terms can be written in an analog way as products of $\vec{S}_{\,\vec{i}}\cdot \vec{S}_{\,\vec{j}}$ due to the SU(2) symmetry of the Hubbard model \cite{Yang10,Reischl04}. 

\begin{figure}
\begin{center}
\includegraphics*[width=\columnwidth]{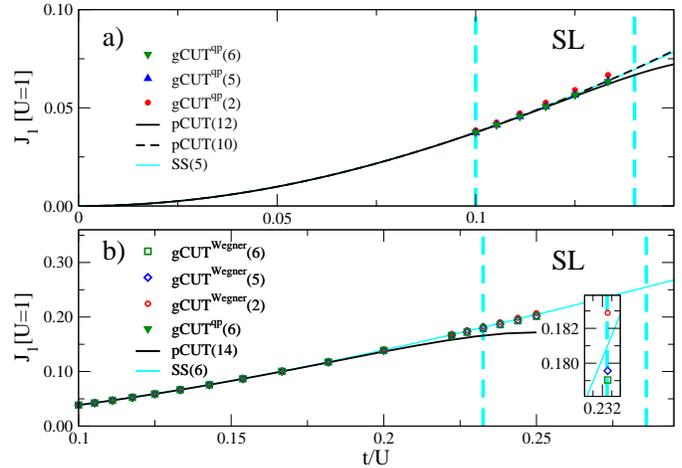}
\end{center}
\caption{(Color online) Nearest-neighbor Heisenberg exchange $J_1$ as a function of $t/U$ for the (a) triangular lattice and (b) the honeycomb lattice. We always compare bare (black lines) and extrapolated order $n$ self-similar approximants of the pCUT series SS($n$) to gCUT curves with varying $n_{\rm max}$ using the qp (filled symbols) and the Wegner generator (open symbols). The first (second) dashed vertical line illustrates for both lattices the beginning (end) of the SL \cite{Yang10,Meng10}.} 
\label{fig:ex2_hubbard_coupling_j1}
\end{figure}

\begin{figure}
\begin{center}
\includegraphics*[width=\columnwidth]{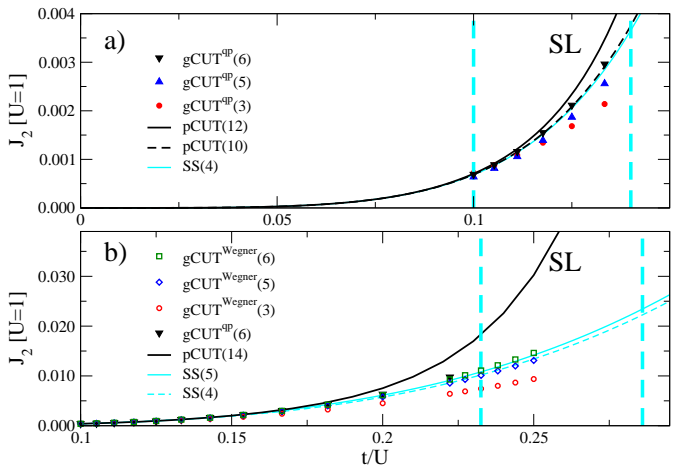}
\end{center}
\caption{(Color online) Next-nearest neighbor Heisenberg exchange $J_2$ as a function of $t/U$ for the (a) triangular lattice and (b) the honeycomb lattice. Notations are equivalent to the ones used in Fig.~\ref{fig:ex2_hubbard_coupling_j1}.} 
\label{fig:ex2_hubbard_coupling_j2}
\end{figure}

In the following we focus on the nearest neighbor and next-nearest neighbor Heisenberg exchanges $J_1$ and $J_2$. We calculated both couplings for the two lattices with gCUTs up to $n_{\rm max}=6$ using the qp and the Wegner generator. As explained above, the generators behave differently if unitarily transformed $q=0$ and $q>0$ states overlap energetically. We have found this to be the case for gCUT(6) around $t/U=0.25$. In this range of parameters one has to use the Wegner generator to obtain an effective spin model. In all other cases we studied the results obtained by both generators are very similar. In total there are $67$ ($11$) different graphs for the triangular (honeycomb) lattice. Additionally, we have performed a pCUT calculation up to order 12 (14) in $t/U$ for the triangular (honeycomb) lattice. The results are shown in Figs.~\ref{fig:ex2_hubbard_coupling_j1}-\ref{fig:ex2_hubbard_coupling_j2}.

We remark that the gCUT curves for different $n_{\rm max}$ smoothly converge in a wide range of $t/U$ parameters. Interestingly, the bare pCUT series become unreliable well before and it is mandatory to use extrapolation schemes. We have found that different approximations like Pad\'e or DlogPad\'e work well for the nearest neighbor exchange $J_1$. In contrast, such standard extrapolation schemes fail for the next-nearest neighbor exchange $J_2$ in the $t/U$ range of the SL on the honeycomb lattice due to the appearance of spurious poles. We therefore have applied so-called self-similar appromixants \cite{Yukalov98}. A self-similar approximant SS$(k)$ of order $k$ is defined as the function
\begin{equation}
   SS(k) = (1+A_1 x)^{n_1} (1+A_2 x)^{n_2} \ldots (1+A_k)^{n_k} \quad , 
\end{equation}
where $A_i$ and $n_i$ are complex coefficients. These coefficients have to be determined by comparing the Taylor expansion of $SS(k)$ in the variable $x$ to the obtained pCUT series with $x=t/U$. We found convincing and convergent results for both couplings on both lattices. The largest deviation between two different subsequent self-similar approximants $SS(k)$ and $SS(k+1)$ are found for $J_2$ on the honeycomb lattice as displayed in Fig.~\ref{fig:ex2_hubbard_coupling_j2}b. Remarkably, the gCUT and the self-similar approximants are in very good agreement for both couplings on both lattices for the range of considered $t/U$ values.

This is a very promising result, since it shows that the gCUT can be at least as good as extrapolations of high-order pCUT series. For the honeycomb lattice, we stress that the recently discovered SL \cite{Meng10} lies not in the purely perturbative regime in contrast to the one on the triangular lattice \cite{Yang10}. One attractive scenario put forward in the literature \cite{Lu10,Clark10} is the relevance of a $J_1$-$J_2$ model for the SL phase on the honeycomb lattice. It is argued that due to the increasing frustration when increasing $t/U$ the long-range ordered N\'eel state found at small values of $t/U$ is unstable towards a SL phase. We find $J_2/J_1\approx 0.06$ at the critical point $U/t=4.3$ obtained from quantum Monte Carlo simulations \cite{Meng10}. It should be clarified in the future whether this rather small ratio is sufficient to destabilize the N\'eel order on the honeycomb lattice or whether other spin-couplings are relevant for the realization of the SL.

%
%

%
%

%
%
%
%
Both examples show that the gCUT is a very robust and flexible tool to derive effective low-energy models in a quantitative fashion. The gCUT is solely limited by the maximal $n_{\max}$ which can still be treated. One therefore demands that the correlation length is not larger than the range of fluctuations captured by the treated graphs. 

If larger $n_{\rm max}$ are needed to capture the physics, one has to develop strategies to treat larger $n$. In practise, this is most likely possible by reducing the number of flow equations such that on the one hand the graph becomes numerically tractable and on the other hand the amplitudes of interest are still quantitatively correct. One therefore has to introduce truncations which are most physically performed like in current sCUT calculations where blocks belonging to a large number of QPs are discarded from the calculation because their influence on the blocks with a few number of QPs is expected to be small. But we note that the current gCUT set up has some intrinsic advantages: i) The existence of a fix point for the full graph is guaranteed by the finite size. One can therefore stop the flow once a convergence of the diagonal couplings of interest is achieved. ii) The quality of these amplitudes can be checked quantitatively by using exact diagonalization for the full model (as long as $n$ is not too large).       

It is one strength of CUTs that also effective observables and dynamical correlation functions can be calculated. To this end any observable has to be transformed with the same unitary transformation as the Hamiltonian. A gCUT for observables amounts therefore to perform the same steps as for the Hamiltonian: i) collect all graphs with fixed $n$. ii) Solve flow equations for the observable on the graphs. iii) Do subcluster substractions and embeddings on the real lattice. 

There are actually many physical problems where gCUT is expected to give important new insights:

1) The pCUT needs an equidistant unperturbed spectrum. The gCUT does not have this limitation. A much larger class of microscopic problems can therefore be tackled including quantum spin models with S$>$1/2, models of mixed degrees of freedom like the t-J model giving rise to unconventional superconductivity, multi-orbital Hubbard models with more than one local energy scale $U$ important for realistic modelling of materials, or bosonic models relevant for cold atoms in optical lattices.

2) The gCUT allows also the calculation of dynamical correlation functions for a much larger class of problems. This opens the possibility to obtain a microscopic understanding of various spectroscopic experiments like it has been done successfully for many pCUT applications. 

3) The gCUT can also be applied in situation where no obvious perturbative limit (as for the discussed examples) can be found. It is only important that an appropriate counting operator $\hat{Q}$ and its eigenbasis can be defined. Note that also symmetry broken phases can be tackled.  

4) The gCUT is especially well suited also in three dimensions since $n_{\rm max}$ and therefore the radius of fluctuations kept in the gCUT is basically constant since only the number of graphs for a given $n_{\rm max}$ increase.  
%
%
%
%

In conclusion, we developped a robust scheme to derive non-perturbatively effective Hamiltonians for strongly correlated quantum lattice models. Such gCUTs combine graph theory and CUTs, and can be applied to any lattice, in any dimension, and for any degree of freedom as long as one sort is gapped. It can be either used to study the properties of elementary excitations of a strongly correlated system or to derive effective low-energy Hamiltonians for infinitely degenerated subspaces. The latter might then be tackled with other powerful techniques like cluster dynamical mean-field theory \cite{Maier05} or functional renormalization group \cite{Salmhofer01,Reuther10}. 

\acknowledgments We thank S.~Dusuel, T.~Fischer, S.~Hamerla, M.~Kamfor, A.~L\"auchli, and G.~S.~Uhrig for fruitful discussion. KPS acknowledges ESF and EuroHorcs for funding through his EURYI.

\end{document}